\documentstyle[sprocl]{article}
\input{psfig}
\def\Journal#1#2#3#4{{#1} {\bf #2}, #3 (#4)}

\def\PRD{{\em Phys. Rev.} D}

\def\be{\begin{equation}}
\def\ee{\end{equation}}
\def\bea{\begin{eqnarray}}
\def\eea{\end{eqnarray}}

\begin{document}
\addtocounter{footnote}{-1}

{\small
\noindent
UNITU-THEP-13/1996  \newline
ANL-PHY-8477-TH-96 \newline
hep-ph/9608471    }

\title{THE STRONG RUNNING COUPLING FROM AN \\
APPROXIMATE GLUON DYSON-SCHWINGER EQUATION$^\dagger $
\footnote{$^\dagger $Supported
by DFG under contract no.\ Al 279/3-1 and by DOE under contract no. 
W-31-109-ENG-38.}}

\author{R. ALKOFER, A. HAUCK}

\address{Institut f\"{u}r Theoretische Physik,
Universit\"{a}t T\"{u}bingen, \\
Auf der Morgenstelle 14, 72076 T\"{u}bingen, Germany}

\author{L. VON SMEKAL}

\address{Physics Division, Argonne National Laboratory,\\
Argonne, Illinois 60439-4843, USA}

\maketitle\abstracts{Using Mandelstam's approximation to the gluon
Dyson-Schwinger equation we calculate the gluon self-energy in a
renormalisation group invariant fashion. We obtain a non-perturbative $\beta $
function. The scaling behaviour near the ultraviolet stable fixed point is in
good agreement with perturbative QCD. No further fixed point for positive
values of the coupling is found: $\alpha_S$ increases without
bound in the infrared. }

The infrared behaviour of the running coupling in strong interactions,
$\alpha_S$, is of great interest, since it may provide an understanding of
confinement. Its study is an intrinsically non-perturbative problem. One
suitable framework to address this problem is provided by the
Dyson-Schwinger equations of QCD. Studies of this infinite tower of
equations rely on specific truncation schemes. Here we will focus on
an approximation to the gluon Dyson-Schwinger equation originally proposed
by Mandelstam.\cite{Man79} This yields a simplified equation for the
inverse gluon propagator in Euclidean momentum space,
\begin{eqnarray}  \label{eq:GluonDSE}
  {D^{-1}}^{\mu\nu}(k)
    = {D_0^{-1}}^{\mu\nu}(k) &+& \frac{g_0^2 N_C}{32\pi^4 k^2}
        \int \!d^4\!q \:  \Gamma_0^{\mu\rho\alpha}(k,q-k,-q)  \\
              && \times D^{\alpha\beta}(q) D_0^{\rho\sigma}(k-q)
                \Gamma_0^{\beta\sigma\nu}(q,k-q,-k)  \quad , \nonumber
\end{eqnarray}
where $D_0$ and $\Gamma_0$ are the bare gluon propagator and the bare
three-gluon vertex. The use of a second bare vertex in the gluon loop is
combined with one of the gluon propagators being bare. This is to account for
some of the dressing of the vertex as entailed by its Slavnov-Taylor
identity.\cite{Man79}

In a manifestly gauge invariant formulation the Dyson--Schwinger equation
for the inverse gluon propagator in the covariant gauge would be transverse
without further adjustments. This may be violated due to the neglect of
ghosts, the violation of Slavnov--Taylor identities and also due to a
regularisation that does not preserve the residual local invariance under 
transformations generated by harmonic gauge functions ($\partial^2
\Lambda(x) =0$). The latter is the case for an $O(4)$ invariant Euclidean
cutoff $\Lambda $, which we will use to regularise eq.~(\ref{eq:GluonDSE}).
Contracting eq. (\ref{eq:GluonDSE}) with the transversal projector,
\begin{equation}
  P^{\mu\nu}(k) = \delta^{\mu\nu} - \frac{k^\mu k^\nu}{k^2} \; .
\end{equation}
leads to an equation for the gluon renormalisation function $G(k^2)$ in
Landau gauge, which after the angular integrations reads\cite{Man79,Atk81}
\bea  \label{eq:Mandelstam}
  \frac{1}{G(k^2)}
     = 1 &+ \frac{g_0^2}{16\pi^2}\frac{1}{k^2} \int_0^{k^2} dq^2
              \left(   \frac{7}{8}\frac{q^4}{k^4}
                     - \frac{25}{4}\frac{q^2}{k^2}
                     - \frac{9}{2} \right) G(q^2)  \\
         &+ \frac{g_0^2}{16\pi^2}\frac{1}{k^2} \int_{k^2}^{\Lambda^2} dq^2
              \left(   \frac{7}{8}\frac{k^4}{q^4}
                     - \frac{25}{4}\frac{k^2}{q^2}
                     - \frac{9}{2} \right) G(q^2) \; . \nonumber
\eea
However, the above equation contains a quadratically ultraviolet divergent
term, which has to be subtracted by a suitable counter term. 
Generally, quadratic ultraviolet divergences can occur only in the part of
the inverse gluon propagator proportional to $\delta^{\mu\nu}$. Therefore,
that part cannot be unambiguously determined, it depends on the routing of
the momenta. This is due to the various violations of gauge invariance
mentioned above. The unambiguous term proportional to $k^\mu k^\nu $ can
be obtained by contracting (\ref{eq:GluonDSE}) with
\begin{equation}
  R^{\mu\nu}(k) = \delta^{\mu\nu} - 4 \, \frac{k^\mu k^\nu}{k^2} \quad .
\end{equation}
In this case, upon angle integration instead of (\ref{eq:Mandelstam}) one
obtains \cite{Bro89}
\bea  \label{eq:Brown}
  \frac{1}{G(k^2)}
     = 1 &+ \frac{g_0^2}{16\pi^2}\frac{1}{k^2} \int_0^{k^2} dq^2
              \left(   \frac{7}{2}\frac{q^4}{k^4}
                     - \frac{17}{2}\frac{q^2}{k^2}
                     - \frac{9}{8} \right) G(q^2)  \\
         &+ \frac{g_0^2}{16\pi^2}\frac{1}{k^2} \int_{k^2}^{\Lambda^2} dq^2
              \left(   \frac{7}{8}\frac{k^4}{q^4}
                     - 7\frac{k^2}{q^2} \right) G(q^2)  . \nonumber
\eea
The logarithmic ultraviolet divergences in eqs. (\ref{eq:Mandelstam})
and (\ref{eq:Brown}) can be removed by multiplicative renormalisation.
Introducing renormalised gluon propagator and coupling the renormalisation
constants $Z_3$ and $Z_g$ are defined by  
$D \to Z_3 D$ and  $g_0 = Z_g g$. The renormalised Dyson--Schwinger
equation for the gluon propagator in Mandelstam approximation then reads,
\begin{equation}  \label{eq:gluon-ren0}
  G(k^2) = \left[ Z_3 +   Z_3^2 Z_g^2 \, \frac{g^2}{16\pi^2}\,
              I_G (k^2) \right]^{-1} \; ,
\end{equation}
with obvious definitions of $I_G$ as a functional of $G$ for the
respective cases ({\it cf.}, eqs. \ref{eq:Mandelstam} and \ref{eq:Brown}).
In Mandelstam approximation one has $Z_g Z_3 =
1$.\cite{us} We adopt a momentum subtraction scheme requiring
the gluon self--energy to vanish at the renormalisation point $\mu$:
$G(\mu^2) = 1$.

\begin{figure}
\centerline{
\psfig{figure=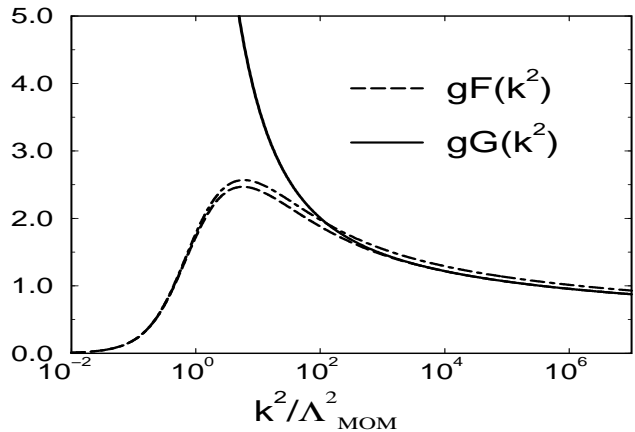,height=6cm,width=10cm}}
 \caption{The gluon renormalisation function $g G(k^2) =8\pi\sigma /k^2 +
gF(k^2)$ for  eq.~\protect\ref{eq:Brown}. The dashed lines
show $g F(k^2)$ for eqs.\ \protect\ref{eq:Brown} and
\protect\ref{eq:Mandelstam}.} 
  \label{fig:gluon}
\end{figure}

The behaviour of $G(k^2)$ for $k^2 \to 0$ can be summarized as follows:
\begin{equation}
gG(k^2) \, =\, {8 \pi \sigma \over k^2  } + g F (k^2) \quad , \qquad g F
(k^2)= a_{00} \left( k^2/\Lambda^2 \right)^\gamma +\ldots \quad .
\end{equation}
Here, $\sigma$ is the string tension, and the subleading term is determined
by\cite{Man79,Atk81}
$$\gamma = \sqrt{\frac{31}{6}} - 1 \approx 1.273, \quad {\rm and}
\quad a_{00}\approx 0.29421$$
for Mandelstam's original equation (\ref{eq:Mandelstam}). For 
eq.\ (\ref{eq:Brown}), we obtain,
$$\gamma  = \frac{2}{9}\sqrt{229} \cos\left( \frac{1}{3}\arccos\left(
      -\frac{1099}{229\sqrt{229}} \right) \right) - \frac{13}{9}
    \approx 1.271\, , \; {\rm and} \quad
a_{00}\approx 0.29446 \; .$$
The numerical solutions to eqs.~(\ref{eq:Brown}) and (\ref{eq:Mandelstam})
for the renormalisation group invariant products $gG$ and $gF$ as functions of
$k^2/\Lambda_{\rm{MOM}}^2$ are shown in fig.~\ref{fig:gluon}. As there are no
qualitative and only little quantitative differences between those two
solutions, we concentrate on the discussion of the solution to
eq.~(\ref{eq:Brown}) in the following.

The non-perturbative and renormalisation group invariant result relates the
string tension $\sigma$ to the QCD scale (the only parameter) by $\sigma = 2
\Lambda^2$. Fixing $\sigma $ to its phenomenological value (it may be
extracted from heavy quarkonium spectra or the slope of Regge trajectories)
gives a value for $\Lambda$ of about 600 MeV. Alternatively, from
$\alpha_S(m_Z^2) \simeq 0.108$ we find $\Lambda \simeq 640$MeV. For a
momentum subtraction scheme with $N_f=0$  this is the correct order of
magnitude.

The scaling behaviour of the solution near the
ultraviolet fixed point is determined by the coefficients $\beta_0 = 14$,
$\beta_1 = 70/3$ and $\gamma_A^0 = 7$ which are reasonably close to the
perturbative values for $N_f = 0$, {\it i.e.},  $\beta_0 = 11$, $\beta_1 = 51
$ and $\gamma_A^0 = 13/2$. 

\begin{figure}
\hskip -.2cm
\parbox{5cm}{
  \psfig{figure=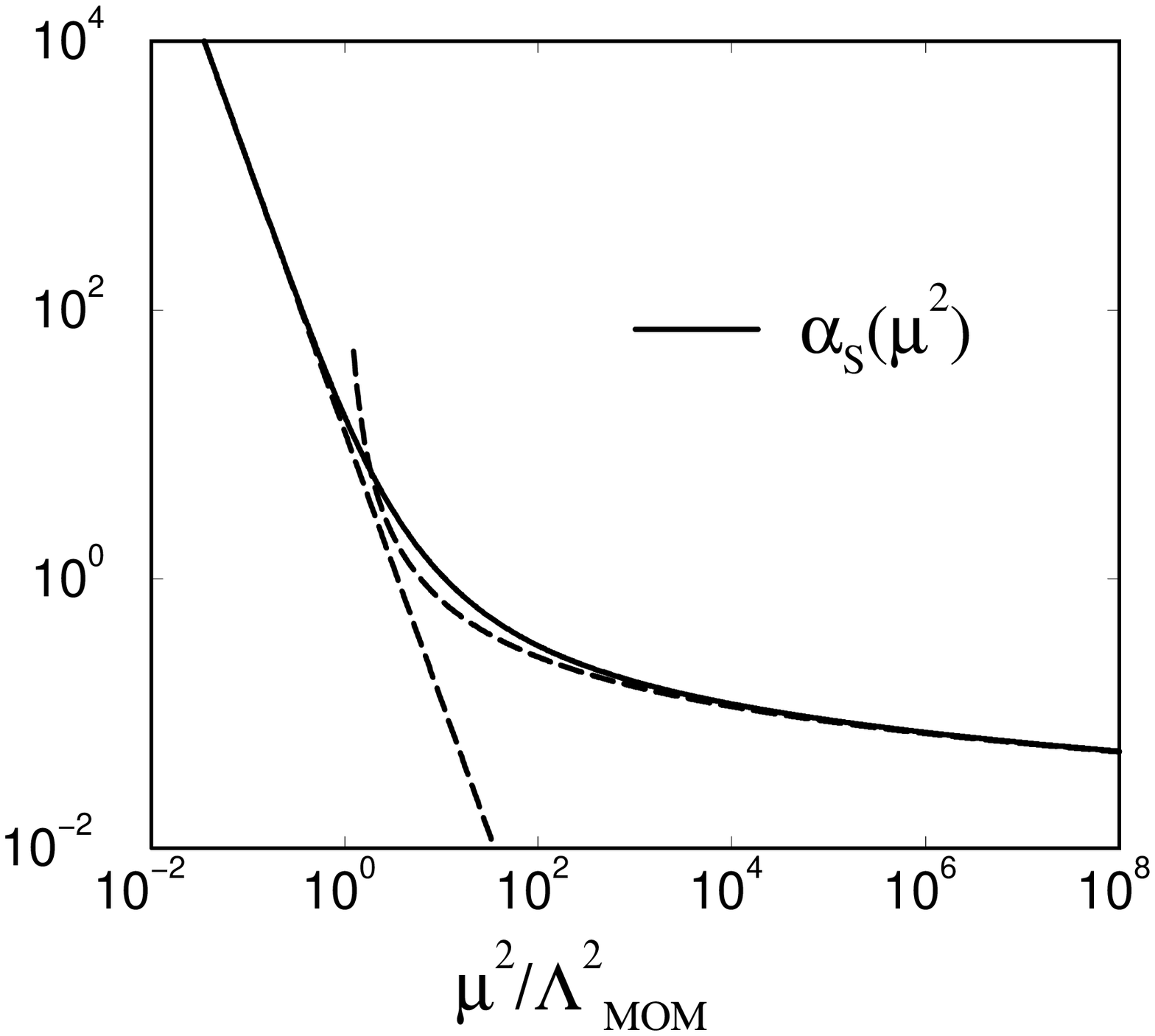,width=6.6cm,rwidth=4cm}}
%\hskip .5cm
\hfill
\parbox{5cm}{
  \psfig{figure=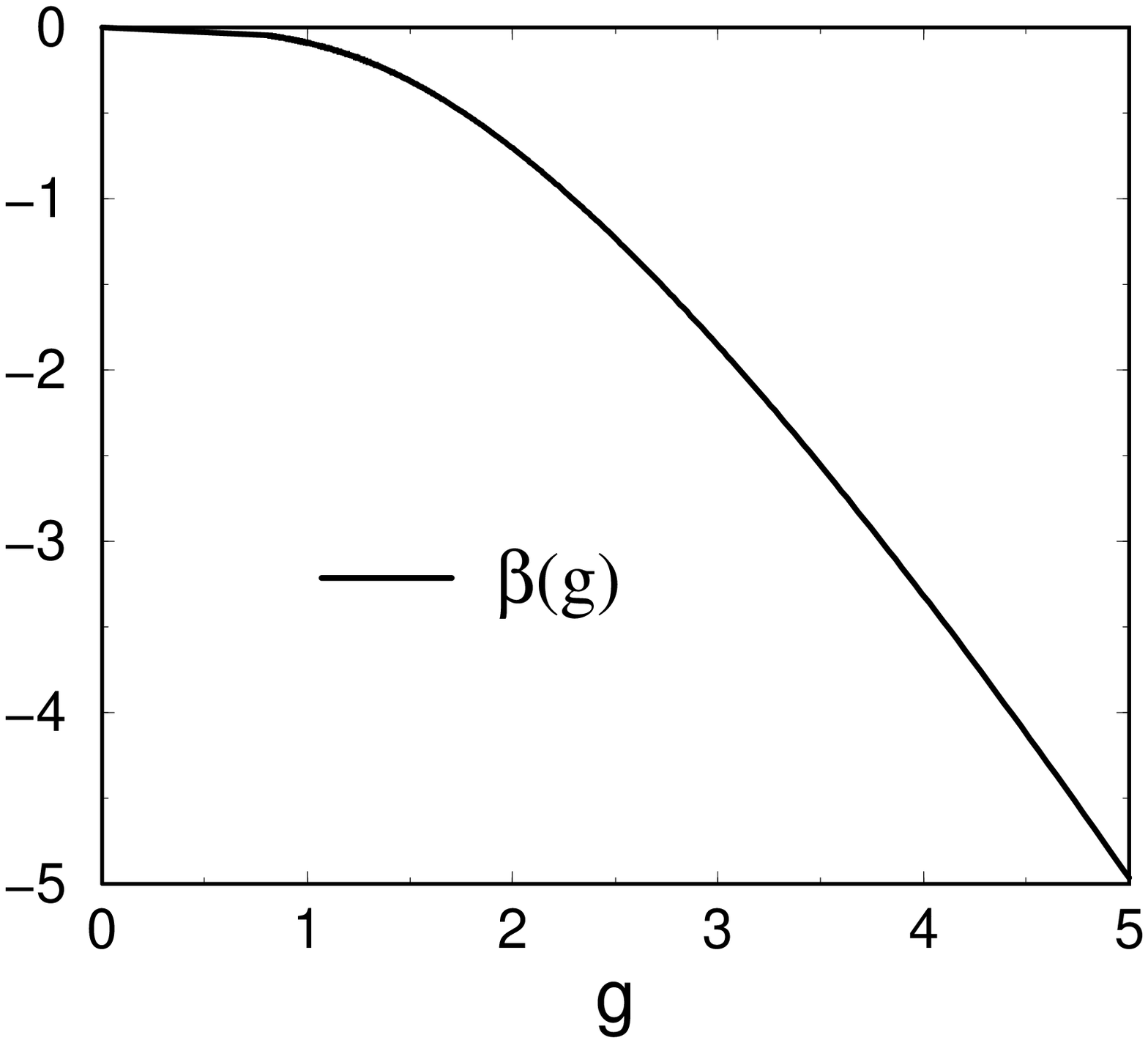,width=6.6cm,rwidth=6cm}}
  \caption{The strong coupling $\alpha_S(\mu^2)$ with its asymptotic limits
and the non-perturbative $\beta$ function of the Mandelstam approximation for
a momentum subtraction scheme.\protect\cite{us}}   
\label{fig:alpha}
\end{figure}

The running coupling, which is obtained for arbitrary scales from the
renormalisation condition, resembles the two-loop perturbative form
asymptotically. 
It is shown together with the corresponding Callan--Symanzik
$\beta$ function in fig.~\ref{fig:alpha}. The running coupling increases
without bound in the infrared, no further fixed point exists.

\section*{References}
\begin{thebibliography}{99}
\bibitem{Man79} S.~Mandelstam, \Journal{\PRD}{20}{3223}{1979}.
\bibitem{Atk81} D. Atkinson, P. W. Johnson, K. Stam, \Journal{\em
J. Math. Phys.}{23}{1917}{1982}. 
\bibitem{Bro89} N.~Brown and M.~R.~Pennington, \Journal{\PRD}{39}{2723}{1989}.
\bibitem{us} A. Hauck, L. v. Smekal, R. Alkofer,
\Journal{\em preprint}{}{hep-ph/9604430}{1996}.

\end{thebibliography}
\end{document}